\documentclass{aastex6}

\begin{document}


\title{Discovery of RR Lyrae Stars in the Nuclear Bulge of the Milky Way}

%% Use \author, \affil, plus the \and command to format author and affiliation 
%% information.  If done correctly the peer review system will be able to
%% automatically put the author and affiliation information from the manuscript
%% and save the corresponding author the trouble of entering it by hand.
%%
%% The \affil should be used to document primary affiliations and the
%% \altaffil should be used for secondary affiliations, titles, or email.

%% Authors with the same affiliation can be grouped in a single
\author{Dante  Minniti\altaffilmark{1,2,3}, Rodrigo Contreras Ramos\altaffilmark{1,4}, Manuela Zoccali\altaffilmark{1,4}, Marina Rejkuba\altaffilmark{5,6}, Oscar A. Gonzalez\altaffilmark{7}, Elena Valenti\altaffilmark{5}, \and Felipe Gran\altaffilmark{1,4}}

\affil{$^1$Instituto Milenio de Astrofisica, Santiago, Chile} 

\affil{$^2$Departamento de Fisica, Facultad de Ciencias Exactas, Universidad Andres Bello\\
Av. Fernandez Concha 700, Las Condes, Santiago, Chile}
\affil{$^3$Vatican Observatory, V00120 Vatican City State, Italy}

\affil{$^4$Pontificia Universidad Catolica de Chile, Instituto de Astrof'sica, Av. Vicuna Mackenna 4860, Santiago, Chile}

\affil{$^5$European Southern Observatory, Karl-Schwarszchild-Str. 2, D85748 Garching bei Muenchen, Germany}

\affil{$^6$Excellence Cluster Universe, Boltzmannstr. 2, 85748, Garching, Germany}

\affil{$^7$UK Astronomy Technology Centre, Royal Observatory, Blackford Hill, Edinburgh, EH9 3HJ, UK}


\altaffiltext{2}{dante@astrofisica.cl}
\altaffiltext{4}{rcontrer@astro.puc.cl}

\begin{abstract}

Galactic nuclei, like the one of the Milky Way, are extreme places with high stellar densities and, in most cases, hosting a supermassive black hole. One of the scenarios proposed for the formation of the Galactic nucleus is by merging of primordial globular clusters \citep{Capuzzo-Dolcetta93}. An implication of this model is that this region should host stars characteristically found in old Milky Way globular clusters. RR Lyrae stars are primary distance indicators, well known representatives of old and metal-poor stellar populations, and therefore regularly found in globular clusters.
Here we report the discovery of a dozen RR Lyrae ab-type stars in the vicinity of the Galactic center, i.e. in the so-called nuclear stellar bulge of the Milky Way. This discovery provides the first direct observational evidence that the Galactic nuclear stellar bulge contains ancient stars ($>$10 Gyr old). Based on this we conclude that merging globular clusters likely contributed to building-up the high stellar density in the nuclear stellar bulge of the Milky Way.
\end{abstract}




\section{Introduction} \label{sec:intro}

There are very limited observational tests that can be applied to shed light on the origins of galactic nuclei, with their stars and black holes. The only galactic nucleus where detailed stellar population properties can be derived with sufficiently high resolution and accuracy is that of the Milky Way, making it therefore a fundamental testbed for different formation models. There we can in principle resolve individual stars, probing a wide range of stellar ages and metallicities. There are two main scenarios proposed for the formation of the nuclear bulge of the Milky Way, and of all galactic nuclei in general: merging of globular clusters  \citep{Tremaine75, Capuzzo-Dolcetta93,  Gnedin14, Guillard16}, and fast gas accretion and star formation onto the central region \citep{Milosavljevic04, Schinnerer08}. 

Here we concentrate on testing the first of these theories. In that scenario, dynamical friction causes orbital decay, dragging globular clusters deep into the potential well, where they merge and form a high density nuclear bulge with a nuclear star cluster at its center.
While merging globular clusters typically bring in old-stellar populations, comparison with observations requires some additional in situ star formation \citep{Antonini15}, or sub-sequent growth of the newly formed nuclear cluster via wet merger with other clusters that bring with them additional gas reservoirs that contribute younger stars \citep{Guillard16}. Therefore, young or intermediate-age stellar populations often dominate the total light in galactic nuclei, even in those cases where they make a minor contribution total the total stellar mass.
It is then very difficult, in an environment like the Galactic center, to establish the presence of the ancient stellar populations, and to estimate their ages and metallicities. Such old populations must be present if the nuclear bulge of the Milky Way was made by merging of primordial globular clusters \citep{Capuzzo-Dolcetta93}. 

Theoretically, if the nuclear stellar bulge formed by merging of several globular clusters, the expected extension of the final merger product is about 100pc \citep{Capuzzo-Dolcetta93, Gnedin14, Antonini12,  Antonini14}. This size appears to be obtained by the simulations regardless the presence or absence of a central massive black hole \citep{Capuzzo-Dolcetta93, Antonini12, Antonini14, Capuzzo-Dolcetta08, Capuzzo-Dolcetta09}. 

Observationally, the nuclear stellar bulge of the Milky Way is well fit by two components: the nuclear star cluster, a compact component with half-light radius of 4 pc (2 arcmin) that dominates the inner ~30 pc, and the nuclear stellar bulge, a shallower component extending out to about 120 pc \citep{Launhardt02}. This size is comparable to the sizes of well studied nuclear stellar bulges of other external galaxies \citep{Hartmann11, Carollo02, Lotz01}.

As globular clusters are tidally disrupted, they yield their stars, including RR Lyrae, to the field. So far, there has been no search for variable stars deep enough to find RR Lyrae in the complex Galactic center region. However, this can now be tested observationally with the VISTA Variables in the Via Lactea (VVV) ESO public survey \citep{Minniti10, Saito12}, that contains deep multi-epoch photometry in the near-IR, allowing to find faint variable sources. 

In the present search we also find numerous bright LPVs/Miras, eclipsing binaries, Cepheids, and microlensing events, which would be reported elsewhere. We concentrate here only on the RR Lyrae because in this context they play a crucial role. Their properties make them prime representatives of the primordial stellar populations of the Milky Way: (a) they have a well known Period-Luminosity relation, and are therefore excellent distance indicators; (b) they have a very narrow range of intrinsic colors that make them excellent reddening indicators; and (c) they are old (age $>$10 Gyr), and metal-poor (i.e. $[Fe/H] < -0.5$).

\section{VVV Survey Photometry} \label{sec:vvv}
The limiting magnitudes ($K_s\sim18$ mag, $J\sim20$ mag) and spatial resolution ($\sim$0.8 arcsec) of the near-IR data provided by the VVV survey enable for the first time a successfully search of RR Lyrae throughout the Galactic center region. 
%We discovered a dozen bonafide RR Lyrae within 36 arcmin (84 pc) of the Galactic center. 
The PSF-fitting photometry of the individual VVV images for $\sim$100 epochs of the tiles b333 and b334 was carried out following the procedure described by \citet{Alonso-Garcia15, Minniti15}. The search for periodic variable stars, phasing of the light curves, and classification of the RR Lyrae type ab were made following the strategies outlined by \citet{Gran16, Dekany13}. We searched for RR Lyrae with magnitudes $12 < Ks < 17$, amplitudes $0.2 < A < 1.0$, periods $0.3 < P < 1.0$\,days.

\begin{figure}[h]
\plotone{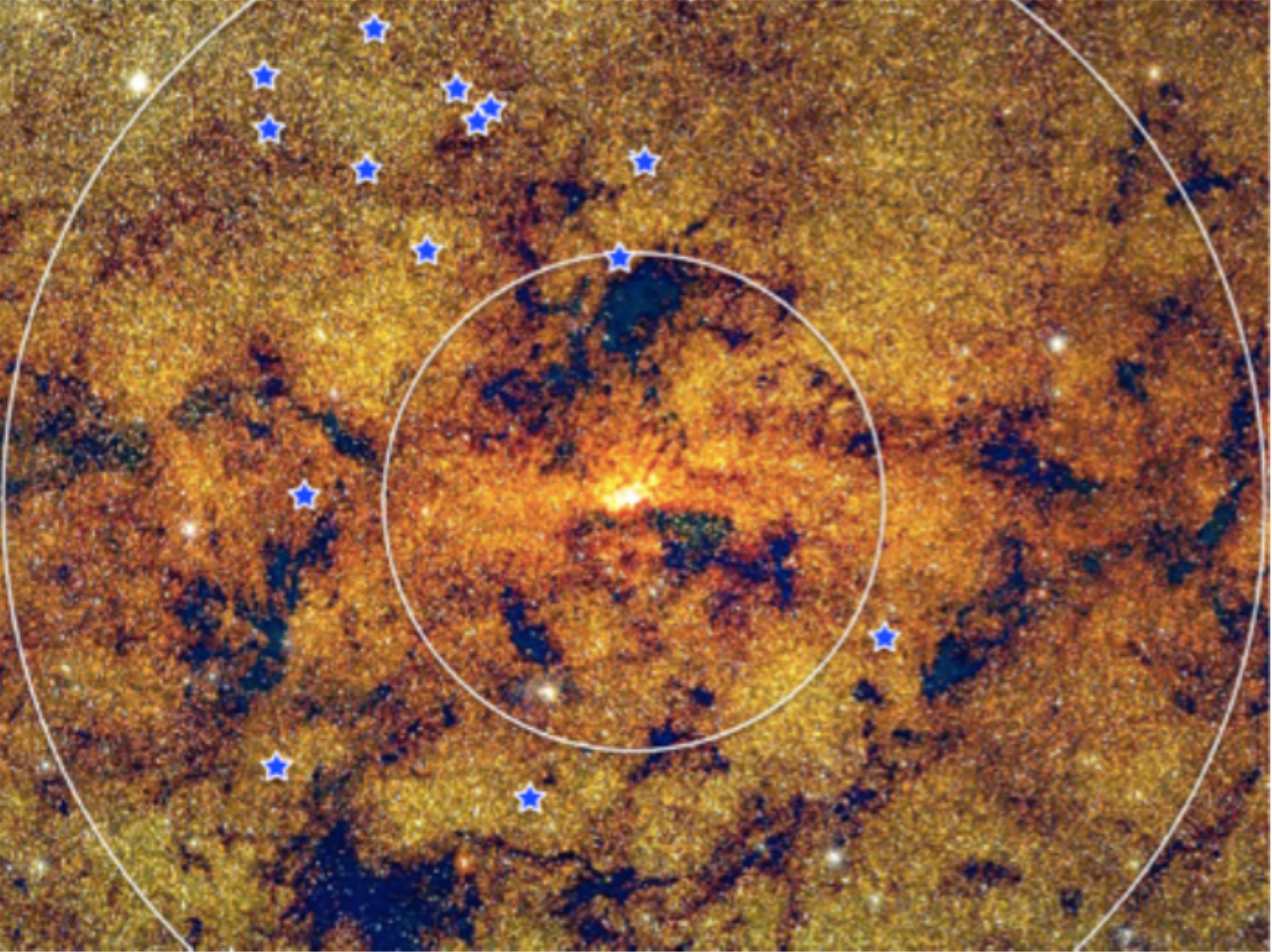}
\caption{Location of the RR Lyrae type ab stars. Near infrared JHKs color image of the Galactic center region from the VVV survey \citep{Minniti10}. The extensions of the nuclear star cluster (R$\sim$35 pc), and the nuclear bulge (R$\sim$100 pc) are indicated with the small and large circles, respectively. One of the leading theories for the formation of the nuclear bulge of the Milky Way is by merging of primordial globular clusters \citep{Launhardt02}. The RR Lyrae from these disrupted clusters are now found throughout this central region (blue stars). One of them (VVV-RRL-40405) has measured distance and reddening consistent with membership to the innermost nuclear star cluster, with a projected distance of 16 arcmin (35 pc) from the Galactic center.
\label{fig:image}}
\end{figure}

\begin{figure}[h]
%\figurenum{2}
\plotone{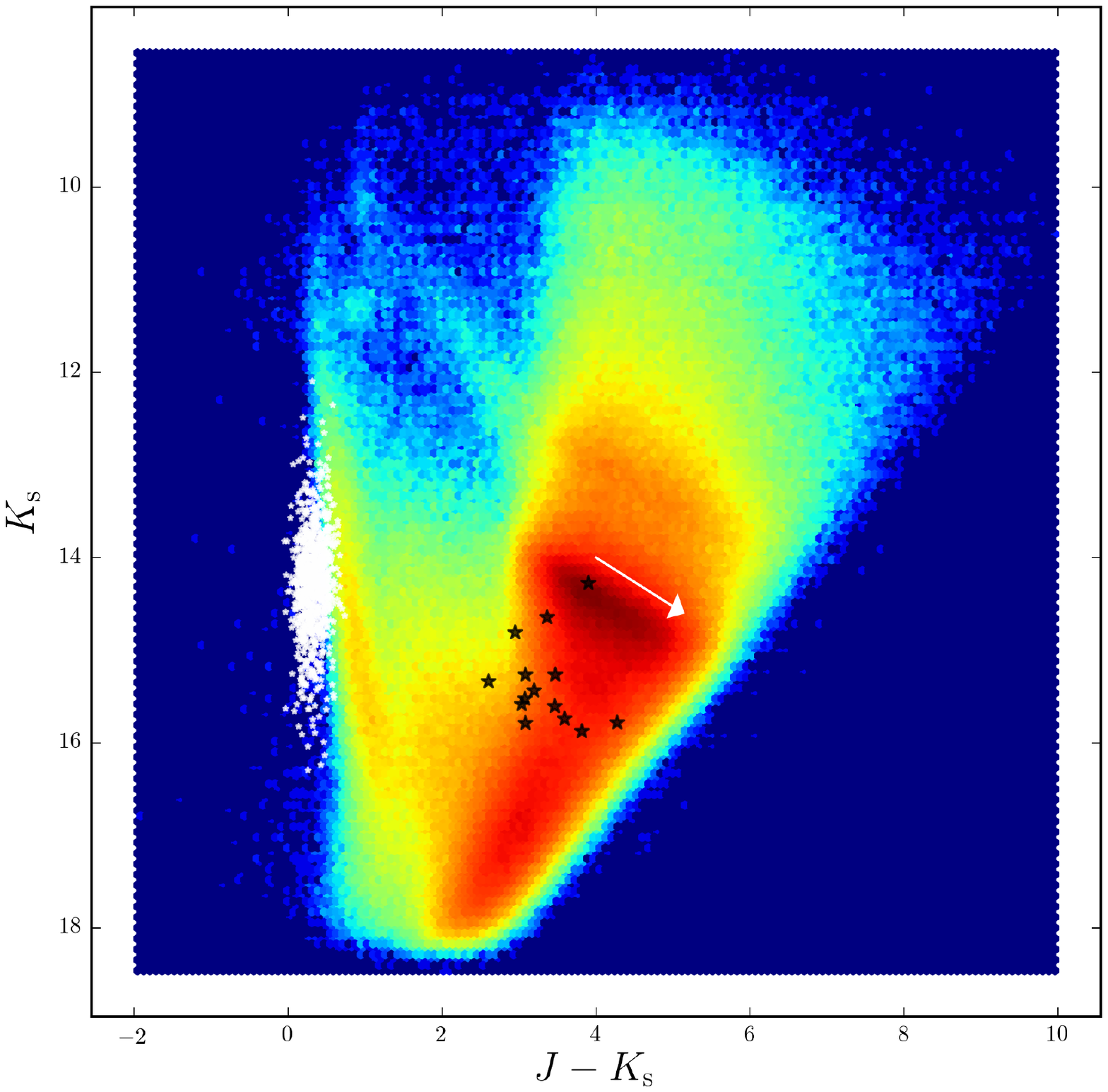}
\caption{
VVV Near-infrared color-magnitude diagram for the Galactic center region showing the position of the nuclear bulge RR Lyrae (black stars). For comparison, the unreddened outer bulge RR Lyrae \citet{Gran16} are shown to the left of the color-magnitude diagram (white dots). The direction of the reddening vector is shown also. This vector is a fit to the shape of the reddened clump giants, and agrees with the slope measured by \citet{Alonso-Garcia15}. We measure the extinctions and distances using the near-IR photometry, which also place these RR Lyrae at the Galactic center region.  
\label{fig:cmd}}
\end{figure}

Extreme crowding and extinction variations are clearly evident in near-IR images taken by the VVV survey (Figure\,\ref{fig:image}). Searching for RR Lyrae in the most crowded and reddened region of the Milky Way is therefore a daunting task, in which several problems need to be faced and sorted out. Specifically, the completeness depends on the position in the field, as these are near-IR mosaics. The stellar density is very high, and the presence of numerous saturated stars in the field that obliterate their surroundings is a limiting factor in the photometry. In addition, the large and differential reddening, highly variable even on small spatial scale, affects the photometric completeness, although the effect is less severe than the crowding. Indeed, the VVV photometry is generally deep enough to reach well below the RR Lyrae region of the color-magnitude diagram even in the most reddened region at the distance of the Galactic center (Figure\,\ref{fig:cmd}). The photometric completeness in this region measured from red clump giants at $14.3 < K_s < 15.9$, the magnitude range of the spanned by the observed RR Lyrae, has an average value of $80\%$ \citep{Valenti16}. However, the variable seeing and uneven sampling of the observed epochs contribute to further reduce the completeness of our sample. The faint magnitudes of the targets also prevent us from finding/classifying RR Lyrae that have very small amplitudes ( $<0.2$ mag).

For all these reasons, we do not claim full completeness for the detection of RR Lyrae, but conversely we expect many more to be found in dedicated high resolution deep searches that might enable to establish the total RR Lyrae density number in the Galactic center region.

The shape of the light curves is also an important limiting factor, with contamination from eclipsing binaries being a serious problem for the sinusoidal light curves, and therefore limiting us to select mostly RRab with asymmetric light curves. The total number of epochs (i.e. points in the light curves) is $\sim$100 epochs, generally sufficient to select RR Lyrae with confidence. However, sampling is a problem, with many good candidates that need to be discarded as aliases. We therefore have many more RR Lyrae candidates for which additional observations are needed in order to measure their periods accurately. These observations would be acquired in the next 3 years as part of the VVV extended survey (VVVX).

\section{The Innermost RR Lyrae} \label{sec:rrl}
\begin{figure}[h]
%\figurenum{2}
\plotone{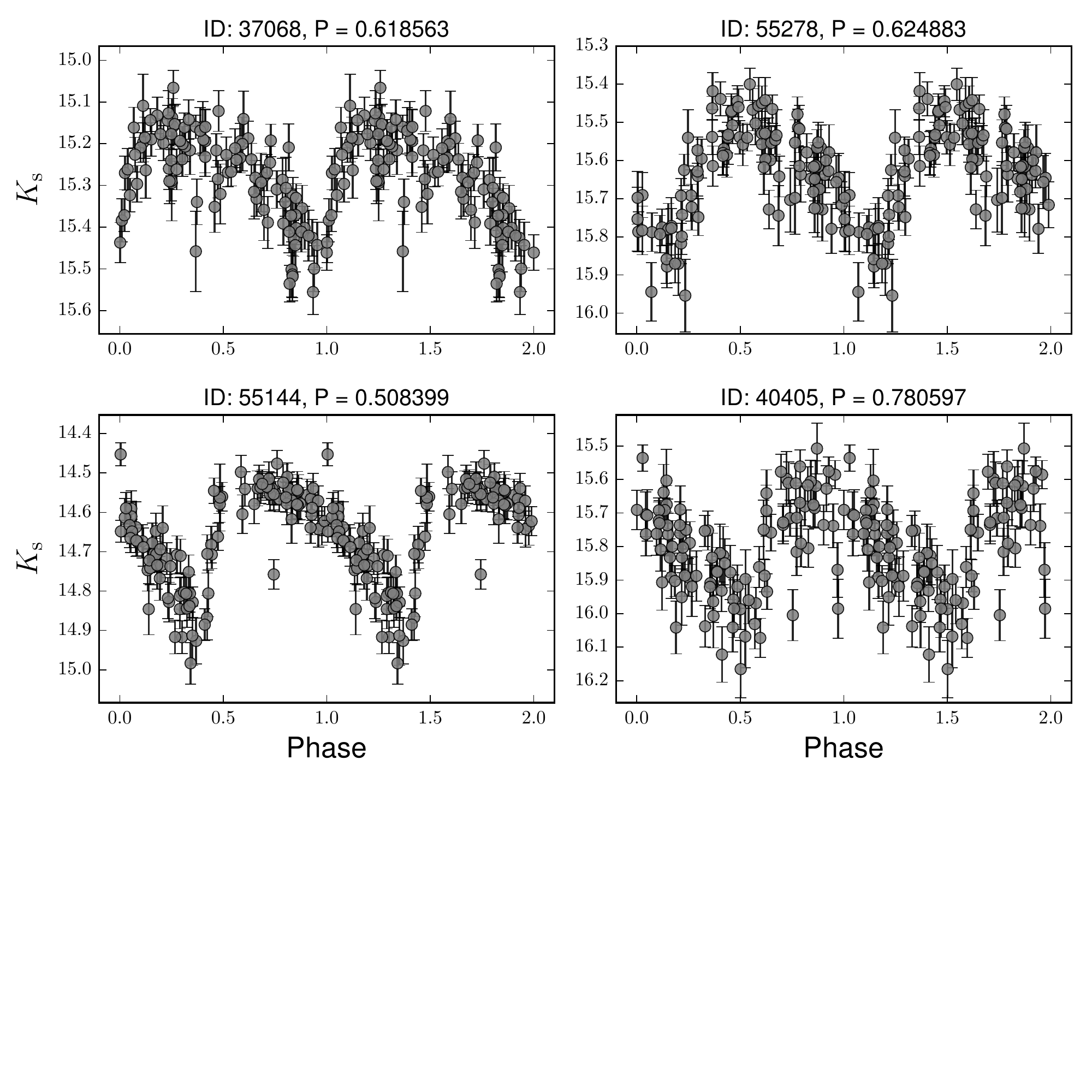}
\caption{Phased light curves for candidate RR Lyrae.
We have concentrated on the search for RR Lyrae type ab stars (funtamental mode pulsators) that have
asymmetric light curves, in order to avoid contamination from short period eclipsing binaries. The presence of these objects allows us to conclude that the nuclear stellar bulge of the Milky Way actually contains a population of RR Lyrae that are very old stars (age $>10$ Gyr).
\label{fig:lcurves}}
\end{figure}

We report here the discovery of a dozen RR Lyrae type ab (fundamental mode pulsators) stars within 36 arcmin (84 pc) from the Galactic center (Figure \ref{fig:image}), plus a couple of c-type RRLyrae (pulsating in their first overtone). We measure accurate positions, projected distances from the Galactic center, mean IR magnitudes and colors, periods, and amplitudes for all of our targets (Table \ref{tab:PhotObs}). The clear variability signature of RR Lyrae, including their characteristic saw-tooth light curve shape (Figure \ref{fig:lcurves}), and their measured amplitudes and periods, are the unambiguous signatures that we are detecting individual ancient and faint RR Lyrae, and not stellar blends or other artefacts.

We also found a candidate type II Cepheid at a projected distance of 45pc (20 arcmin) from the Galactic center. This type II Cepheid with P= 1.809 days, mean $K_s=14.66$, and $(J-K_s)= 4.02$, is also representative of an old and metal-poor population present in the vicinity of the nuclear star cluster. A fundamental implication about the old age of the variable stars found here is that the nuclear stellar bulge must have been in place since the origins of the Milky Way. In addition, we discovered several more bonafide RR Lyrae over a wider area, within 36 - 50 arcmin (84 - 109 pc) of the Galactic center, just outside of the nuclear bulge.
 
The near-IR color-magnitude diagram obtained from PSF fitting photometry shows the high extinction of the field where these RR Lyrae have been discovered (Figure \ref{fig:cmd}). The target RR Lyrae are fainter and bluer than the bulge red clump giants. When taking into account the large extinction difference across the bulge, the comparison of the color-magnitude diagram of these RR Lyrae and in other regions of the bulge \citep{Minniti10, Saito12} is consistent with them being RR Lyrae located in the region of the Galactic center.

\subsection{Extinction corrections} \label{sec:ext}
 
Extinction corrections to the measured near-IR magnitudes is a mandatory step in order to assess the location of the sample RR Lyrae within the Galactic nucleus. The RR Lyrae lie in the instability strip, which is a narrow band in the color-magnitude diagram, and their intrinsic colors can be assumed to be $(J-K)_0=0.15 \pm 0.05$ mag. Although we only have $K_s$-band light curves and a single (or a few) J-band epoch, the color corrections for de-reddened RR Lyrae due to the single J-band observation is negligible \citep{Gran16, Dekany13}. In fact, the color variation along the light curves is typically small in the near-IR ($\Delta(J-K_s)<0.05$ mag).
 When computing the reddening for a specific target, the most important systematic error is the uncertain slope of the reddening law \citep{Gonzalez12, Nataf16, Majaess16}. 
 For example, comparing $A_k=0.528E(J-K)$ given by \citet{Nishiyama09} with $A_k=0.72E(J-K)$ from \citet{Cardelli89}, and $A_k=0.435E(J-K)$ from \citet{Alonso-Garcia15}, the corresponding
differences for the typical extinction values of the Galactic center region ($E(J-K) \sim$ 3.0 to 4.0) are significant. Adopting the most recent value given by  \citet{Alonso-Garcia15} that applies to the VVV data, and that also agrees with the slope of the reddened red giant clump seen in the color-magnitude diagram (Figure \ref{fig:cmd}), we find for each candidate the reddening $E(J-K_s)$ and extinction $A_{k}$ listed in Table\,\ref{tab:StelParam}.

\subsection{Metallicities} \label{sec:met}
In order to explore the properties of globular clusters that could have initially formed the Galactic nuclear bulge, we examine the properties of the Oosterhoff types I and II globular cluster populations \citep{Oosterhoff39, Catelan09}. A way to distinguish between these two populations is by measuring the average periods of their RR Lyrae, which are shorter in the mean for type I ($<P>=0.55$ days) than for type II Oosterhoff cluster populations ($<P>=0.65$ days) \citep{Catelan09}. We find that the distribution of periods of the nuclear bulge RR Lyrae has a mean of $P=0.55$ days, resembling an Oosterhoff type I population (more metal-rich than $[Fe/H]= -1.6$ dex). 
%The derived mean period of the sample RR Lyrae is also similar to the population of bulge globular clusters.

Alternatively, the mean metallicities for RR Lyrae type ab can be estimated using their period-amplitude-metallicity relation. After discarding the RR Lyrae type c, we obtain mean metallicities $<[Fe/H]>=$-1.0, -1.4, and -1.3 dex for the sample RR Lyrae using the calibrations from \citet{Alcock00, Yang10, Feast10}, respectively. 

The Bailey diagram is shown in Figure\,\ref{fig:bailey}, in comparison with the bulge RR Lyrae. The resulting individual metallicities using the Galactic RR Lyrae calibration from \citet{Feast10} (that should only be taken as indicative until spectroscopic measurements become available), are listed in Table\,\ref{tab:StelParam}. Therefore, we suggest that most of the merged clusters were Oosterhoff type I globulars (with $-1.6 <[Fe/H]< -0.9$ dex). Even though we cannot discard that a few of the primordial globular clusters that merged into the nuclear stellar bulge might have been very metal-poor (with $[Fe/H]< -2$\,dex like VVV-RRL-40405), they do not appear to be the dominant population. Interestingly, the star VVV-RRL-40405 that is located only 16 arcmin away from the Galactic center is the most metal-poor star of this sample, with $[Fe/H] \sim -2.2$\,dex.  
This very metal-poor object is the first RR Lyrae that is a likely member of the nuclear star cluster. Knowing its intrinsic color, distance, and extinction (Table\,\ref{tab:StelParam}), we can estimate its visual magnitude, V = 29.5 mag, much too faint and out of reach for current optical instruments.

\subsection{Distances} \label{sec:dist}
In order to compute distances we also have to take into account the different sources of errors. An important systematic error is the absolute magnitude scale for the period-luminosity (P-L) relation. We make two different assumptions to compute distances, in order to illustrate the
uncertainties involved. First (case 1), we use the P-L relation given by eq.\,14 from \citet{Muraveva15} that is
based on the cleanest sample of Hipparcos and HST RR Lyrae parallaxes, and the extinctions listed in Table\,\ref{tab:StelParam}, obtaining distances that are consistent with membership to the Galactic nuclear bulge. Second (case 2), as the P-L relation also depends on the chemical composition (the so-called P\--L\--Z relation), assuming that these RR Lyrae are the debris of Oosterhoff type I globular clusters, we adopt a mean $[Fe/H]= -1.0$ dex in the P\--L\--Z relation of \citet{Alonso-Garcia15}, as well as a steeper extinction law from \citet{Nishiyama09}. In this way we also obtain distances that are consistent with membership to the Galactic nuclear bulge, but larger in the mean by about 500 pc that in the first case considered above. For both cases, Table\,\ref{tab:StelParam} summarizes the RR Lyrae reddenings, extinctions, distance moduli and distances in kpc from the Sun, and metallicities. In all this we have also assumed that the photometric VVV zero point error is negligible ( $<0.01$ mag).

The distance distributions of the RR Lyrae in our sample compared with 1019 RR Lyrae type ab recently discovered in the outer bulge from the VVV survey \citep{Gran16} have consistent peak values. The distribution of the present sample is very concentrated, much more so than the observed distribution of RR Lyrae in the inner and outer bulge. Considering all these uncertainties, we can conclude that most of our RR Lyrae are located at the distance of the Galactic center. Only two of the brightest ones (VVV-RRL-65743, and VVV-RRL-55144) could be foreground objects, although we cannot completely discard the possibility that they are blended sources. The two main types of sources that brighten an object in this region would be bulge clump stars (which are redder), or foreground disk stars (which are bluer). However, the colors alone cannot help to distinguish these possibilities given the large and non-uniform reddening.

\section{Conclusions} \label{sec:con}
For the first time, we find that there are RR Lyrae in the region well within the nuclear stellar bulge of our Galaxy, suggesting that they could be the remains of the primordial globular clusters
that built up the nuclear bulge. The dozen RR Lyrae stars presented here give a limit to the age and metallicity of the nuclear bulge, and thus provide valuable clues about its origin. While there is ample evidence that the stellar population of the nuclear star cluster is composite, containing a mixture of young, intermediate and old stellar populations \citep{Genzel10, Schoedel14, Chatzopoulos15}, the RR Lyrae stars we found suggest that the nuclear bulge is very  old ($>10$ Gyr), perhaps as old as the Milky Way itself.

\begin{figure}[h]
%\figurenum{2}
\plotone{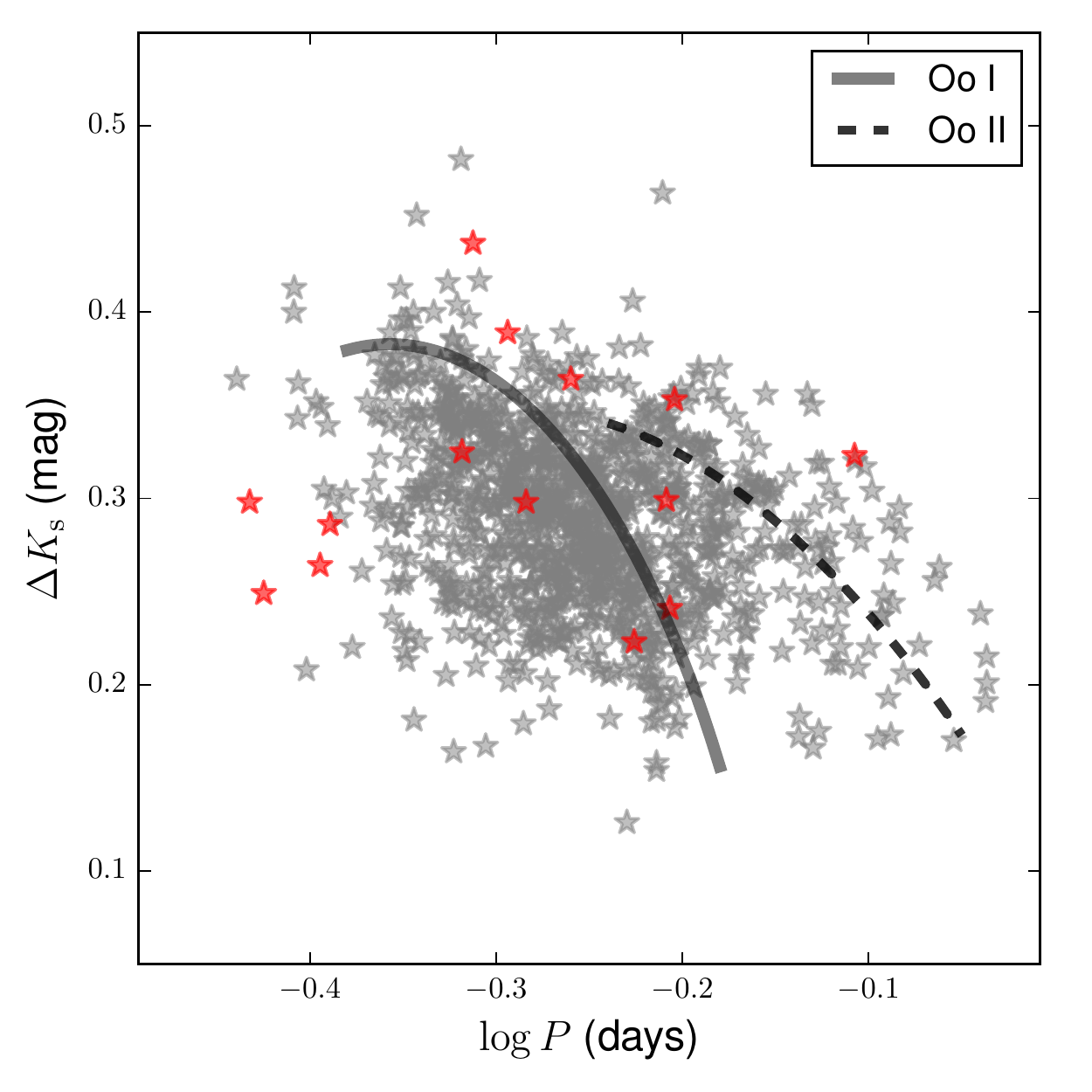}
\caption{Bailey diagram: amplitude vs period for the nuclear bulge RR Lyrae (red stars) compared with 1019 RR Lyrae type ab found in the outer bulge \citet{Gran16}. The main ridge lines for the Oosterhoff types I and II are indicated, which are the left and right groups, respectively. Most of the nuclear bulge RR Lyrae share the location of the more metal-rich Oosterhoff type I population that has a metallicity $-0.9 < [Fe/H]< -1.6$ dex \citep{Catelan09}.
\label{fig:bailey}}
\end{figure}

Are these RR Lyrae special in any way? How do they compare with the RR Lyrae previously found in the Milky Way bulge? RR Lyrae are numerous in globular clusters and in the Milky Way halo, and are taken as prime tracers of old ($>$10 Gyr), and metal-poor stellar populations. However, not all globular cluster RR Lyrae are similar. For example, there are two populations of globular clusters \citep{Oosterhoff39, Catelan09}: the Oosterhoff type I clusters, that are more metal-rich ($-0.9<[Fe/ H]<-1.6$ dex), and the Oosterhoff type II clusters, that are more metal-poor ($[Fe/H]< -1.6$ dex). With a mean period of $<P>=0.55$ days, most of the RR Lyrae discovered here are representative of an Oosterhoff type I population (Figure\,\ref{fig:bailey}). 

Overall, the properties of the present sample are consistent with the bulge RR Lyrae population \citep{Gran16}, being more concentrated to the Galactic centre. The evidence supports the scenario where the nuclear stellar bulge was originally made out of a few globular clusters that merged through dynamical friction \citep{Capuzzo-Dolcetta93, Guillard16}, and as such it could well be the most massive and oldest surviving star cluster of our Galaxy.

%%%%%%%%%%%%%%%%%%%%%   TABLES %%%%%%%%%%%%%%%%%%%%%%%%%%%%%%%

\begin{deluxetable}{ccccccccc}
\tablecaption{Photometric Observations \label{tab:PhotObs}}
\tablehead{
\colhead{ID} & \colhead{RA} & \colhead{DEC} & \colhead{R} & \colhead{K$_s$} & \colhead{J} & \colhead{$(J-K_s)$} & \colhead{P} & \colhead{A}\\
\colhead{} & \colhead{(J2000)} & \colhead{(J2000)} & \colhead{arcmin} & \colhead{} & \colhead{} & \colhead{} & \colhead{days} & \colhead{}
}
%\colnumbers
\startdata
40405 & 266.17164326 & -28.86342987 & 15.5 & 15.78 & 20.05 & 4.27 & 0.780597 & 0.32 \\
65743 & 266.40425763 & -29.31187570 & 18.3 & 14.27 & 18.17 & 3.90 & 0.594484 & 0.22 \\
55278 & 266.79651367 & -29.08202592 & 20.4 & 15.60 & 19.07 & 3.47 & 0.624883 & 0.35 \\
37068 & 266.28880383 & -28.68220478 & 20.7 & 15.26 & 18.34&  2.98 & 0.618563 & 0.30 \\
80042 & 266.62151461 & -28.71032145 & 20.8 & 15.44 & 18.63 & 3.19 & 0.408153 & 0.29 \\
84844 & 266.06131561 & -28.83296846 & 21.4 & 15.87 & 19.69&  3.82 & 0.549641 & 0.36 \\
58214 & 266.11652879 & -28.65279182 & 26.5 & 15.79 & 18.87 & 3.08 & 0.403153 & 0.26 \\
55144 & 266.10448919 & -28.65918014 & 26.6 & 14.64 & 18.00 & 3.36 & 0.508399 & 0.39 \\
89691 & 266.24177968 & -28.58633606 & 26.9 & 15.74 & 19.33 & 3.59 & 0.376027 & 0.25 \\
33007 & 266.92238653 & -28.83231159 & 28.6 & 15.27 & 18.74 & 3.47 & 0.621276 & 0.24 \\
42332 & 266.10220084 & -28.61729278 & 28.7 & 15.34 & 17.93 & 2.59 & 0.520099 & 0.30 \\
65271 & 266.26347962 & -28.47343076 & 33.1 & 14.81 & 17.75 & 2.94 & 0.369534 & 0.30 \\
8444 & 266.09247811 & -28.51255523 & 34.3 & 15.52 & 18.60 & 3.08 & 0.487035 & 0.44 \\
33289 & 266.20795127 & -28.43858475 & 35.9 & 15.58 & 18.61 & 3.03 & 0.480476 & 0.33  \\
\enddata
\tablenotetext{a}{Typical photometric errors are $\sigma_{Ks} = 0.01$ mag, and $\sigma_{J} = 0.06$ mag. Periods are good to $10^{-5}$ days, and $K_s$-band amplitude errors are $\sigma_A=0.02$ mag.}
\end{deluxetable}

\begin{deluxetable}{ccccccccc}
\tablecaption{Measured stellar parameters \label{tab:StelParam}}
\tablehead{
\colhead{ID} & \colhead{$E(J-K_s)$} & \colhead{$A_K$} & \colhead{$(m-M)_0$}\tablenotemark{a} & \colhead{D$_{\odot}$}\tablenotemark{a} & \colhead{$(m-M)_0$}\tablenotemark{b} & \colhead{D$_{\odot}$}\tablenotemark{b} & \colhead{$[Fe/H]$} & \colhead{Comments}\\
\colhead{} & \colhead{} & \colhead{} & \colhead{} & \colhead{kpc} & \colhead{} & \colhead{kpc} & \colhead{dex} & \colhead{}
}
%\colnumbers
\startdata
40405 & 4.12 & 1.79 & 14.67 & 8.6 &14.92 & 9.6 & -2.2 & RRab \\
65743 & 3.75 & 1.63 &13.02 & 4.0 & 13.27 & 4.5 & -1.5 & RRab foreground?\\
55278 & 3.32 & 1.44 &14.59 & 8.3 & 14.85 & 9.3 & -1.6 & RRab \\
37068 & 2.83 & 1.23 &14.45 & 7.8 & 14.67 & 8.6 & -1.6  & RRab \\
80042 & 3.04 & 1.32 &14.09 & 6.6 & 14.32 & 7.3 & -0.6 & RRab \\
84844 & 3.67 & 1.60 & 14.56 & 8.2 & 14.82 & 9.2  &-1.3 &  RRab \\
58214 & 2.93 & 1.27 & 14.47 & 7.8 & 14.71 & 8.8 & -0.6 & RRab \\
55144 & 3.21 & 1.40 & 13.45 & 4.9 & 13.70 & 5.5 & -1.2 & RRab foreground?\\
89691 & 3.44 & 1.50 & 14.12 & 6.7 & 14.36 & 7.4 & -0.5 & RRc \\
33007 & 3.32 & 1.44 & 14.26 & 7.1 & 14.50 & 7.9  & -1.6 & RRab \\
42332 & 2.44 & 1.06 & 14.51 & 8.0 & 14.75 & 8.9 & -1.2 &  RRab \\
65271 & 2.79 & 1.21 & 13.54 & 5.1 & 13.69 & 5.8 & -0.4 & RRc \\
 8444 & 2.93  & 1.27 & 14.41 & 7.6 & 14.66 & 8.6 & -1.1 & RRab \\
33289 & 2.88 & 1.25 & 14.47 & 7.8 & 14.72 & 8.8 & -1.0 & RRab\\
\enddata
\tablenotetext{a}{{\it Case 1}: By using the P\--L relation given by eq.\,14 from \citet{Muraveva15} and the extinctions listed in Table\,\ref{tab:PhotObs}}
\tablenotetext{b}{{\it Case 2}: By using the P\--L\--Z relation of \citet{Alonso-Garcia15} and assuming a mean $[Fe/H]=-1$\,dex}
\tablenotetext{c}{Typical reddening errors are $\sigma_{E(J-Ks)}=0.1$ mag, and metallicity errors are $\sigma_{[Fe/H]}=0.3$ dex.}
\end{deluxetable}

%%%%%%%%%%%%%%%%%%%%%   FIGURES %%%%%%%%%%%%%%%%%%%%%%%%%%%%%%%

%\begin{figure}[h]
%%\figurenum{2}
%\plotone{fig5.pdf}{angle=0}
%\caption{RR Lyrae distance distributions. Case 1 (left), and case 2 (right) distance distributions for RR Lyrae in the %Galactic center region compared to those of the outer bulge from \citet{Gran16} (shown in red and blue, respectively). The %distances for these samples were computed using the VVV near-IR photometry, and the distributions have been normalised to %their respective peak number counts. The RR Lyrae in the Galactic nuclear bulge region seem to be more concentrated than %the ones of the outer bulge. Even though the absolute distances are uncertain due to the different reddening slopes and %period-luminosity relations, the high concentration demonstrates that the majority of them are located in the Galactic %center region, and are not foreground or background RR Lyrae.  
%\label{fig:f5}}
%\end{figure}

\acknowledgments

We gratefully acknowledge the use of data from the VVV ESO Public Survey program ID 179.B-2002 taken with the VISTA telescope, and data products from the Cambridge Astronomical Survey Unit (CASU). The VVV Survey data are made public at the ESO Archive. Support for the authors is provided by the BASAL Center for Astrophysics and Associated Technologies (CATA) through grant PFB-06, and the Ministry for the Economy, Development, and Tourism, Programa Iniciativa Cientifica Milenio through grant IC120009, awarded to the Millennium Institute of Astrophysics (MAS). We acknowledge support from FONDECYT Regular grants No. 1130196 (D.M.), and No. 1150345 (M.Z. and F.G.). We are also grateful to the Aspen Center for Physics where our work was supported by National Science Foundation grant PHY-1066293, and by a grant from the Simons Foundation (D.M. and M.Z.).


%% for the paper.  Each keyword is check against the master list during
%% copy editing.  Individual instruments can be provided in parentheses,
%% after the keyword, but they are not verified.

\vspace{5mm}
\facilities{ESO, VIRCAM@VISTA}

\bibliographystyle{aasjournal}
%\bibliography{RRLbiblio}

\begin{thebibliography}{}
\expandafter\ifx\csname natexlab\endcsname\relax\def\natexlab#1{#1}\fi

\bibitem[{{Alcock} {et~al.}(2000){Alcock}, {Allsman}, {Alves}, {Axelrod},
  {Basu}, {Becker}, {Bennett}, {Cook}, {Drake}, {Freeman}, {Geha}, {Griest},
  {King}, {Lehner}, {Marshall}, {Minniti}, {Nelson}, {Peterson}, {Popowski},
  {Pratt}, {Quinn}, {Stubbs}, {Sutherland}, {Tomaney}, {Vandehei}, \&
  {Welch}}]{Alcock00}
{Alcock}, C., {Allsman}, R.~A., {Alves}, D.~R., {et~al.} 2000, \aj, 119, 2194

\bibitem[{{Alonso-Garc{\'{\i}}a} {et~al.}(2015){Alonso-Garc{\'{\i}}a},
  {D{\'e}k{\'a}ny}, {Catelan}, {Contreras Ramos}, {Gran}, {Amigo}, {Leyton}, \&
  {Minniti}}]{Alonso-Garcia15}
{Alonso-Garc{\'{\i}}a}, J., {D{\'e}k{\'a}ny}, I., {Catelan}, M., {et~al.} 2015,
  \aj, 149, 99

\bibitem[{{Antonini}(2014)}]{Antonini14}
{Antonini}, F. 2014, \apj, 794, 106

\bibitem[{{Antonini} {et~al.}(2015){Antonini}, {Barausse}, \&
  {Silk}}]{Antonini15}
{Antonini}, F., {Barausse}, E., \& {Silk}, J. 2015, \apj, 812, 72

\bibitem[{{Antonini} {et~al.}(2012){Antonini}, {Capuzzo-Dolcetta},
  {Mastrobuono-Battisti}, \& {Merritt}}]{Antonini12}
{Antonini}, F., {Capuzzo-Dolcetta}, R., {Mastrobuono-Battisti}, A., \&
  {Merritt}, D. 2012, \apj, 750, 111

\bibitem[{{Capuzzo-Dolcetta}(1993)}]{Capuzzo-Dolcetta93}
{Capuzzo-Dolcetta}, R. 1993, \apj, 415, 616

\bibitem[{{Capuzzo-Dolcetta} \&
  {Mastrobuono-Battisti}(2009)}]{Capuzzo-Dolcetta09}
{Capuzzo-Dolcetta}, R., \& {Mastrobuono-Battisti}, A. 2009, \aap, 507, 183

\bibitem[{{Capuzzo-Dolcetta} \& {Miocchi}(2008)}]{Capuzzo-Dolcetta08}
{Capuzzo-Dolcetta}, R., \& {Miocchi}, P. 2008, \mnras, 388, L69

\bibitem[{{Cardelli} {et~al.}(1989){Cardelli}, {Clayton}, \&
  {Mathis}}]{Cardelli89}
{Cardelli}, J.~A., {Clayton}, G.~C., \& {Mathis}, J.~S. 1989, \apj, 345, 245

\bibitem[{{Carollo} {et~al.}(2002){Carollo}, {Stiavelli}, {Seigar}, {de Zeeuw},
  \& {Dejonghe}}]{Carollo02}
{Carollo}, C.~M., {Stiavelli}, M., {Seigar}, M., {de Zeeuw}, P.~T., \&
  {Dejonghe}, H. 2002, \aj, 123, 159

\bibitem[{{Catelan}(2009)}]{Catelan09}
{Catelan}, M. 2009, \apss, 320, 261

\bibitem[{{Chatzopoulos} {et~al.}(2015){Chatzopoulos}, {Fritz}, {Gerhard},
  {Gillessen}, {Wegg}, {Genzel}, \& {Pfuhl}}]{Chatzopoulos15}
{Chatzopoulos}, S., {Fritz}, T.~K., {Gerhard}, O., {et~al.} 2015, \mnras, 447,
  948

\bibitem[{{D{\'e}k{\'a}ny} {et~al.}(2013){D{\'e}k{\'a}ny}, {Minniti},
  {Catelan}, {Zoccali}, {Saito}, {Hempel}, \& {Gonzalez}}]{Dekany13}
{D{\'e}k{\'a}ny}, I., {Minniti}, D., {Catelan}, M., {et~al.} 2013, \apjl, 776,
  L19

\bibitem[{{Feast} {et~al.}(2010){Feast}, {Abedigamba}, \&
  {Whitelock}}]{Feast10}
{Feast}, M.~W., {Abedigamba}, O.~P., \& {Whitelock}, P.~A. 2010, \mnras, 408,
  L76

\bibitem[{{Genzel} {et~al.}(2010){Genzel}, {Eisenhauer}, \&
  {Gillessen}}]{Genzel10}
{Genzel}, R., {Eisenhauer}, F., \& {Gillessen}, S. 2010, Reviews of Modern
  Physics, 82, 3121

\bibitem[{{Gnedin} {et~al.}(2014){Gnedin}, {Ostriker}, \&
  {Tremaine}}]{Gnedin14}
{Gnedin}, O.~Y., {Ostriker}, J.~P., \& {Tremaine}, S. 2014, \apj, 785, 71

\bibitem[{{Gonzalez} {et~al.}(2012){Gonzalez}, {Rejkuba}, {Zoccali}, {Valenti},
  {Minniti}, {Schultheis}, {Tobar}, \& {Chen}}]{Gonzalez12}
{Gonzalez}, O.~A., {Rejkuba}, M., {Zoccali}, M., {et~al.} 2012, \aap, 543, A13

\bibitem[{{Gran} {et~al.}(2016){Gran}, {Minniti}, {Saito}, {Zoccali},
  {Gonzalez}, {Navarrete}, {Catelan}, {Contreras Ramos}, {Elorrieta},
  {Eyheramendy}, \& {Jord{\'a}n}}]{Gran16}
{Gran}, F., {Minniti}, D., {Saito}, R.~K., {et~al.} 2016, \aap, 591, A145

\bibitem[{{Guillard} {et~al.}(2016){Guillard}, {Emsellem}, \&
  {Renaud}}]{Guillard16}
{Guillard}, N., {Emsellem}, E., \& {Renaud}, F. 2016, \mnras, 461, 3620

\bibitem[{{Hartmann} {et~al.}(2011){Hartmann}, {Debattista}, {Seth},
  {Cappellari}, \& {Quinn}}]{Hartmann11}
{Hartmann}, M., {Debattista}, V.~P., {Seth}, A., {Cappellari}, M., \& {Quinn},
  T.~R. 2011, \mnras, 418, 2697

\bibitem[{{Launhardt} {et~al.}(2002){Launhardt}, {Zylka}, \&
  {Mezger}}]{Launhardt02}
{Launhardt}, R., {Zylka}, R., \& {Mezger}, P.~G. 2002, \aap, 384, 112

\bibitem[{{Lotz} {et~al.}(2001){Lotz}, {Telford}, {Ferguson}, {Miller},
  {Stiavelli}, \& {Mack}}]{Lotz01}
{Lotz}, J.~M., {Telford}, R., {Ferguson}, H.~C., {et~al.} 2001, \apj, 552, 572

\bibitem[{{Majaess} {et~al.}(2016){Majaess}, {Turner}, {Dekany}, {Minniti}, \&
  {Gieren}}]{Majaess16}
{Majaess}, D., {Turner}, D., {Dekany}, I., {Minniti}, D., \& {Gieren}, W. 2016,
  ArXiv e-prints, arXiv:1607.08623

\bibitem[{{Milosavljevi{\'c}}(2004)}]{Milosavljevic04}
{Milosavljevi{\'c}}, M. 2004, \apjl, 605, L13

\bibitem[{{Minniti} {et~al.}(2010){Minniti}, {Lucas}, {Emerson}, {Saito},
  {Hempel}, {Pietrukowicz}, {Ahumada}, {Alonso}, {Alonso-Garcia}, {Arias},
  {Bandyopadhyay}, {Barb{\'a}}, {Barbuy}, {Bedin}, {Bica}, {Borissova},
  {Bronfman}, {Carraro}, {Catelan}, {Clari{\'a}}, {Cross}, {de Grijs},
  {D{\'e}k{\'a}ny}, {Drew}, {Fari{\~n}a}, {Feinstein}, {Fern{\'a}ndez
  Laj{\'u}s}, {Gamen}, {Geisler}, {Gieren}, {Goldman}, {Gonzalez}, {Gunthardt},
  {Gurovich}, {Hambly}, {Irwin}, {Ivanov}, {Jord{\'a}n}, {Kerins}, {Kinemuchi},
  {Kurtev}, {L{\'o}pez-Corredoira}, {Maccarone}, {Masetti}, {Merlo},
  {Messineo}, {Mirabel}, {Monaco}, {Morelli}, {Padilla}, {Palma}, {Parisi},
  {Pignata}, {Rejkuba}, {Roman-Lopes}, {Sale}, {Schreiber}, {Schr{\"o}der},
  {Smith}, {}, {Soto}, {Tamura}, {Tappert}, {Thompson}, {Toledo}, {Zoccali}, \&
  {Pietrzynski}}]{Minniti10}
{Minniti}, D., {Lucas}, P.~W., {Emerson}, J.~P., {et~al.} 2010, \na, 15, 433

\bibitem[{{Minniti} {et~al.}(2015){Minniti}, {Contreras Ramos},
  {Alonso-Garc{\'{\i}}a}, {Anguita}, {Catelan}, {Gran}, {Motta}, {Muro},
  {Rojas}, \& {Saito}}]{Minniti15}
{Minniti}, D., {Contreras Ramos}, R., {Alonso-Garc{\'{\i}}a}, J., {et~al.}
  2015, \apjl, 810, L20

\bibitem[{{Muraveva} {et~al.}(2015){Muraveva}, {Palmer}, {Clementini}, {Luri},
  {Cioni}, {Moretti}, {Marconi}, {Ripepi}, \& {Rubele}}]{Muraveva15}
{Muraveva}, T., {Palmer}, M., {Clementini}, G., {et~al.} 2015, \apj, 807, 127

\bibitem[{{Nataf} {et~al.}(2016){Nataf}, {Gonzalez}, {Casagrande}, {Zasowski},
  {Wegg}, {Wolf}, {Kunder}, {Alonso-Garcia}, {Minniti}, {Rejkuba}, {Saito},
  {Valenti}, {Zoccali}, {Poleski}, {Pietrzy{\'n}ski}, {Skowron},
  {Soszy{\'n}ski}, {Szyma{\'n}ski}, {Udalski}, {Ulaczyk}, \&
  {Wyrzykowski}}]{Nataf16}
{Nataf}, D.~M., {Gonzalez}, O.~A., {Casagrande}, L., {et~al.} 2016, \mnras,
  456, 2692

\bibitem[{{Nishiyama} {et~al.}(2009){Nishiyama}, {Tamura}, {Hatano}, {Kato},
  {Tanab{\'e}}, {Sugitani}, \& {Nagata}}]{Nishiyama09}
{Nishiyama}, S., {Tamura}, M., {Hatano}, H., {et~al.} 2009, \apj, 696, 1407

\bibitem[{{Oosterhoff}(1939)}]{Oosterhoff39}
{Oosterhoff}, P.~T. 1939, The Observatory, 62, 104

\bibitem[{{Saito} {et~al.}(2012){Saito}, {Hempel}, {Minniti}, {Lucas},
  {Rejkuba}, {Toledo}, {Gonzalez}, {Alonso-Garc{\'{\i}}a}, {Irwin},
  {Gonzalez-Solares}, {Hodgkin}, {Lewis}, {Cross}, {Ivanov}, {Kerins},
  {Emerson}, {Soto}, {Am{\^o}res}, {Gurovich}, {D{\'e}k{\'a}ny}, {Angeloni},
  {Beamin}, {Catelan}, {Padilla}, {Zoccali}, {Pietrukowicz}, {Moni Bidin},
  {Mauro}, {Geisler}, {Folkes}, {Sale}, {Borissova}, {Kurtev}, {Ahumada},
  {Alonso}, {Adamson}, {Arias}, {Bandyopadhyay}, {Barb{\'a}}, {Barbuy},
  {Baume}, {Bedin}, {Bellini}, {Benjamin}, {Bica}, {Bonatto}, {Bronfman},
  {Carraro}, {Chen{\`e}}, {Clari{\'a}}, {Clarke}, {Contreras}, {Corvill{\'o}n},
  {de Grijs}, {Dias}, {Drew}, {Fari{\~n}a}, {Feinstein},
  {Fern{\'a}ndez-Laj{\'u}s}, {Gamen}, {Gieren}, {Goldman},
  {Gonz{\'a}lez-Fern{\'a}ndez}, {Grand}, {Gunthardt}, {Hambly}, {Hanson},
  {He{\l}miniak}, {Hoare}, {Huckvale}, {Jord{\'a}n}, {Kinemuchi}, {Longmore},
  {L{\'o}pez-Corredoira}, {Maccarone}, {Majaess}, {Mart{\'{\i}}n}, {Masetti},
  {Mennickent}, {Mirabel}, {Monaco}, {Morelli}, {Motta}, {Palma}, {Parisi},
  {Parker}, {Pe{\~n}aloza}, {Pietrzy{\'n}ski}, {Pignata}, {Popescu}, {Read},
  {Rojas}, {Roman-Lopes}, {Ruiz}, {Saviane}, {Schreiber}, {Schr{\"o}der},
  {Sharma}, {Smith}, {Sodr{\'e}}, {Stead}, {Stephens}, {Tamura}, {Tappert},
  {Thompson}, {Valenti}, {Vanzi}, {Walton}, {Weidmann}, \&
  {Zijlstra}}]{Saito12}
{Saito}, R.~K., {Hempel}, M., {Minniti}, D., {et~al.} 2012, \aap, 537, A107

\bibitem[{{Schinnerer} {et~al.}(2008){Schinnerer}, {B{\"o}ker}, {Meier}, \&
  {Calzetti}}]{Schinnerer08}
{Schinnerer}, E., {B{\"o}ker}, T., {Meier}, D.~S., \& {Calzetti}, D. 2008,
  \apjl, 684, L21

\bibitem[{{Sch{\"o}del} {et~al.}(2014){Sch{\"o}del}, {Feldmeier}, {Neumayer},
  {Meyer}, \& {Yelda}}]{Schoedel14}
{Sch{\"o}del}, R., {Feldmeier}, A., {Neumayer}, N., {Meyer}, L., \& {Yelda}, S.
  2014, Classical and Quantum Gravity, 31, 244007

\bibitem[{{Tremaine} {et~al.}(1975){Tremaine}, {Ostriker}, \&
  {Spitzer}}]{Tremaine75}
{Tremaine}, S.~D., {Ostriker}, J.~P., \& {Spitzer}, Jr., L. 1975, \apj, 196,
  407

\bibitem[{{Valenti} {et~al.}(2016){Valenti}, {Zoccali}, {Gonzalez}, {Minniti},
  {Alonso-Garc{\'{\i}}a}, {Marchetti}, {Hempel}, {Renzini}, \&
  {Rejkuba}}]{Valenti16}
{Valenti}, E., {Zoccali}, M., {Gonzalez}, O.~A., {et~al.} 2016, \aap, 587, L6

\bibitem[{{Yang} {et~al.}(2010){Yang}, {Sarajedini}, {Holtzman}, \&
  {Garnett}}]{Yang10}
{Yang}, S.-C., {Sarajedini}, A., {Holtzman}, J.~A., \& {Garnett}, D.~R. 2010,
  \apj, 724, 799

\end{thebibliography}


\listofchanges

\end{document}